# Gaplessness and the Coulomb anomaly in the strongly disordered films of Molybdenum Carbide

P. Kulkarni[1], P. Szabo[2], M. Zemlicka[3], M. Grajcar[4]

*Centre of Ultra Low Temperature, Institute of Experimental Physics, Slovak Academy of Sciences and P.J. Safarik University, SK-04001 Kosice, Slovakia*
*Department of Experimental Physics, Comenius University, SK-84248 Bratislava, Slovakia and Institute of Physics, Slovak Academy of Sciences, Bratislava, Slovakia*
*prasanna1609@gmail.com*

**Abstract.** Gaplessness was observed in the disordered films of MoC on approaching to the superconductor to insulator transition by reducing the film thickness. The gaplessness is attributed to the enhanced Coulomb interactions due to the loss of screening in the presence of strong disorder in the films.

## INTRODUCTION

The superconductor to insulator transition (SIT) in the thin films can be achieved by varying the thickness, magnetic field or disorder [1-3]. In the presence of strong disorder the competition between the attractive pairing interaction and the enhanced Coulomb interactions due to loss of screening governs the characteristic superconducting state in the films. The microscopic drive for the destruction of the superconductivity needs investigation using the local probes. Here using the low temperature Scanning tunneling microscopy, we confirm the fermionic scenario of the SIT in the thin MoC films, which are uniform in the grain sizes. We also present the gaplessness which occurs as a result of continuously vanishing superconducting pairing amplitude in the films as the SIT is approached by reducing the thickness.

## EXPERIMENTAL

Thin films of MoC were prepared by using the magnetron reactive sputtering technique. The target of highly pure Molybdenum (Mo) was sputtered onto the sapphire substrates in an argon acetylene atmosphere. The structurally well characterized films were investigated for the local conductance properties using a homemade scanning tunneling microscope operating at 400 mK and with the possibility to apply a magnetic field upto 8 T. The MoC films were mounted along with a pure gold substrate, which is used to clean the Au tip in-situ at low temperatures.

## RESULTS AND DISCUSSIONS

Figure 1 shows the thickness dependence of the superconducting transition temperature ($T_c$) in the MoC films. Inset in Figure 1 shows the temperature dependence of the square resistance for three different thicknesses, t = 3, 5 and 10 nm of the MoC films. In all the three cases the superconducting transition remains relatively sharp till SIT and the near zero values of the square resistance were achieved in the superconducting state.

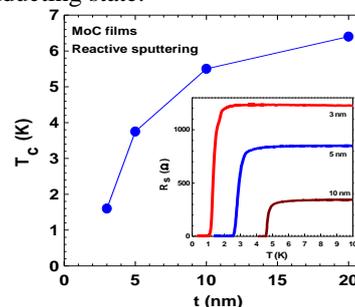

**Figure 1.** The thickness versus the superconducting transition temperature in thin MoC films. Inset shows the temperature variation of the square resistance.

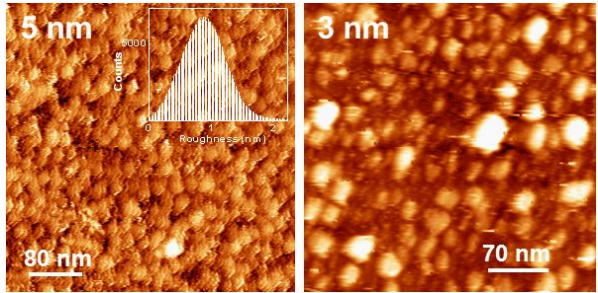

**Figure 2.** The topographic images of the MoC films with 5 nm and 3 nm thicknesses. The inset in the left panel shows the surface roughness in the form of a histogram.

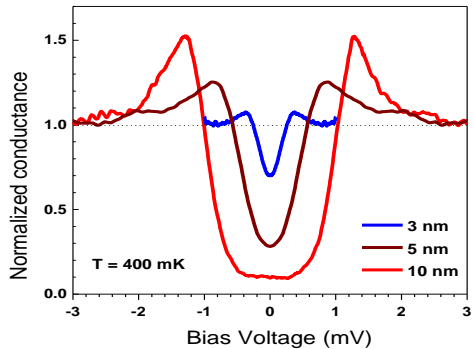

**Figure 3.** The superconducting gap as a function of the MoC film thickness.

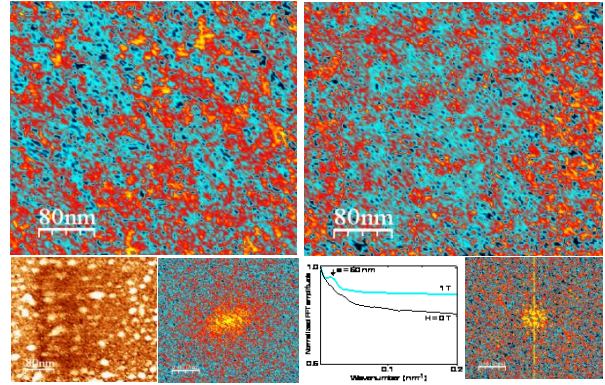

**Figure 4.** The spectral maps in 5 nm MoC film at 0 T and 1 T.

Figure 2 shows the topographic images recorded at 400 mK at zero magnetic field. The inset in the left panel shows the surface roughness of the 5 nm film. The uniform crystallites are visible on the surface of the film and are relatively densely packed in both the films. The roughness of both the films is less than 1 nm and only has variation of the particle heights. No major structural damages are seen on the films suggesting the tunneling characteristics are directly measured from the clean MoC surfaces.

Figure 3 shows the tunneling spectra measured in three films, with t = 3, 5 and 10 nm at 400 mK in the zero magnetic field. For a comparison all three curves are plotted together after normalizing the conductance at higher bias to unity in each case. In 10 nm film, the finite states are observed near zero bias suggesting the gaplessness in the presence of disorder. This feature dramatically enhances as the film thickness reduces to 5 nm and then down to 3 nm. The quasiparticle peaks in each case are clearly observed in the tunneling spectra. This is possibly the first systematic observation of the closing of the superconducting energy gap as a function of the film thickness which characterizes the disorder in the MoC films. The magnitude of the superconducting gap proportionally decreases following with the transition temperature.

In Figure 4, the panel in the left corner shows the topographic image in the 5 nm film over an area of 400 nm x 400 nm. The local I-V curves, total 128 x 128, were measured over this area in the bias voltage range from -3 mV to +3 mV and were normalized at 3 mV on both sides. The conductance values, total 128 x 128, were selected at $V_{bias}$ = 0.4 mV and plotted spatially over the area of 400 nm x 400 nm. The resulting spectral map is shown in the top left panel. The changes in the contrast show the conductance fluctuations in the film. The correlation among these fluctuations can be demonstrated by taking an FFT of the spectral map (the second lower panel from left). The bright cloud around the central region describes the extent of the correlations in terms of the inverse distance. In the graph, the third lower panel, the curve at H = 0 T shows a monotonically varying correlation length for the long range conductance fluctuations in the film. The top panel on the right contains the conductance map after the application of the magnetic field of 1 T on the same region of 400 nm x 400 nm of the film. The normalized conductance curves, total 128 x 128, produces a conductance map at 0.4 mV which shows a relatively periodic changes in the contrast. The FFT image, at the right lower corner, displays a torus like shape around the central region. The corresponding correlation length displays a peak at around 50 nm in the graph. If we consider the inter-vortex spacing at H = 1 T, it is around 50 nm, which matches with the position of the peak in the graph.

In figure 5 the magnetic field dependence of the tunneling conductance curves is shown for t = 3 nm MoC film. As the magnetic field is increased to 3 T the quasiparticle peaks are suppressed and the spectra broaden significantly. At 6 T a strongly reduced tunneling conductance curve was seen and the dependence of conductance is linear on the logarithmic scale of the bias voltage as shown in the inset.

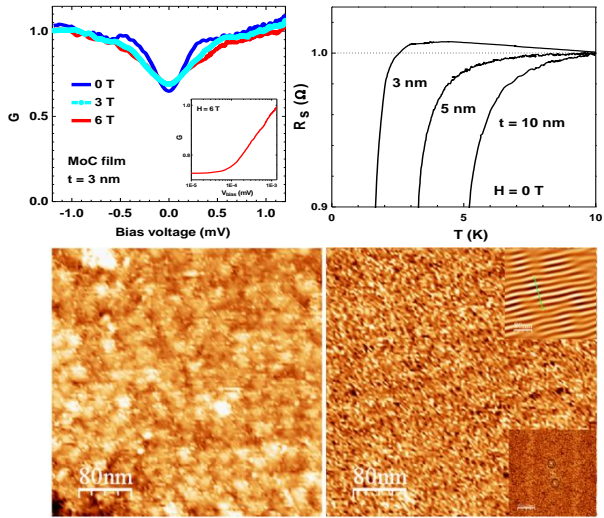

**Figure 5.** The left panel on top shows the superconducting gaps as a function of applied magnetic field in 3 nm MoC film. The right panel shows the normalized transport curves in all the three films. Lower panels show the topographic image on the left and the spectral map on the right at 400 mK and 0 T.

This shows that the normal state (above $H_c = 5$ T) has the reduced density of states which follows a logarithmic dependence. The panel on the right shows the square resistance of the films on the normalized scale at 10 K. Clearly the square resistance increases in the disordered metallic state till the rise is intercepted by the arrival of the superconducting transition.

The panel at the left lower corner the topographic image in $t = 3$ nm film is shown at 400 mK. Uniform particle sizes are seen on the surface of the film over an area 400 nm x 400 nm. On this area, we measured the tunneling IV curves, total 128 x 128, and the conductance at 0.4 mV bias voltage was plotted spatially. The spectral map is shown in the panel on right. The periodic changes of the tunneling conductance are seen in the spectral map extending over distances much larger than the crystalline sizes. The FFT image of the spectral map, shown in the lower inset, has two bright spots which describes the periodicity in the modulations. The filtered image with the conductance modulations are shown in the inset (ii). The periodicity along the green line in the image shows the value around 30 nm.

Our results in MoC films highlight two important features of the disordered superconducting films. The Coulomb interactions in the disordered metallic state of the films play a crucial role in deciding the spatial distribution of the tunneling conductance in the superconducting state. The conductance fluctuations are continuous and long range in the disordered superconducting films. In 5 nm MoC film, the superconducting state co-existing with the conductance fluctuations is reorganized with the application of the magnetic field and the tendency towards the formation of the vortices was seen. This indicates that the superconducting phase tends to remain intact and only affected by the magnetic field while the superconducting pairing interactions weaken at the spatial positions in the film. In the strongly disordered film with $t = 3$ nm, the uptrend in the square resistance shows the insulating like behavior in the disordered metallic state. Such behavior was previously observed in the TiN films and the tunneling spectra showed a strong suppression near zero bias in the normal state following the description of the Altshuler-Aronov zero bias anomaly [4-6]. The gaplessness in the MoC films is attributed to the competition between the superconducting pairing correlations and the enhanced Coulomb repulsive interactions due to the loss of screening in the presence of strong disorder. The later was asserted by the magnetic field above the critical value in $t = 3$ nm film. The inelastic scattering processes sets up the charge density modulations and at these positions the density of states also spatially varies [6]. In MoC films the superconducting gap is likely to close at the SIT and the strongly disordered normal state will restore which has no signatures of the superconducting correlations across SIT. New observation in our measurements is the gaplessness observed in the local tunneling spectra. This feature simply mimics the general character of the superconducting state when the temperature and the magnetic field destroy the superconducting state. Further studies need to explore the structure of the superconducting density of states in the disordered films.

## ACKNOWLEDGMENTS

P. Kulkarni acknowledges the financial support from COST action MP1201. Authors acknowledge the help and the discussions with P. Samuely, J. Kacmarchik and J. Gabriel.